\documentclass[12pt]{article}

\usepackage{amsmath,amssymb,amsfonts}
\usepackage{graphicx,epsfig}
\usepackage{titlesec} %Title format
\usepackage{verbatim}
\usepackage{syntonly}
\titleformat{\section}
  {\large\bfseries}{\thesection.}{0.5em}{}
\titleformat{\subsection}
  {\normalfont\bfseries}{\thesubsection.}{0.5em}{}

\numberwithin{equation}{section}

\makeatletter

\renewcommand{\fnum@figure}{\textbf{Fig.~\thefigure}}
\renewcommand{\fnum@table}{\textbf{Table~\thetable}}

\makeatother

\begin{document}

%%%%% Definitions and newcommands%%%%%
\def\al{\alpha}
\def\be{\beta}
\def\ga{\gamma} \def\Ga{\Gamma}
\def\de{\delta} \def\De{\Delta}
\def\ep{\epsilon}
\def\et{\eta}
\def\th{\theta} \def\Th{\Theta}
\def\ka{\kappa}
\def\la{\lambda} \def\La{\Lambda}
\def\si{\sigma} \def\Si{\Sigma}
\def\ph{\phi} \def\Ph{\Phi}
\def\ch{\chi}
\def\om{\omega} \def\Om{\Omega}

\def\ba{{\bar a}}
\def\bb{{\bar b}}
\def\bc{{\bar c}}
\def\bd{{\bar d}}
\def\bi{{\bar \imath}}
\def\bj{{\bar \jmath}}
\def\bz{\bar z}

\def\bga{{\bar \gamma}}
\def\bka{{\bar \kappa}}
\def\bth{{\bar \theta}}
\def\bla{{\bar \lambda}}
\def\bch{{\bar \chi}}
\def\bOm{{\bar \Omega}}
\def\etd{\eta^\dagger}
\def\chd{\chi^\dagger}

\def\tc{\widetilde c}
\def\tG{\widetilde G}
\def\tZ{\widetilde Z}
\def\tY{\widetilde Y}
\def\tH{\widetilde H}
\def\tGa{\widetilde \Gamma}
\def\cH{\mathcal H}
\def\tcH{\widetilde{\mathcal H}}
\def\cF{\mathcal F}
\def\cL{\mathcal L}
\def\cM{\mathcal M}
\def\cP{\mathcal P}

\def\hA{{\widehat A}}
\def\hB{{\widehat B}}
\def\hC{{\widehat C}}
\def\hM{{\widehat M}}
\def\hN{{\widehat N}}
\def\hP{{\widehat P}}
\def\hQ{{\widehat Q}}
\def\ha{{\widehat a}}
\def\hb{{\widehat b}}
\def\hc{{\widehat c}}
\def\hi{{\widehat \imath}}
\def\hj{{\widehat \jmath}}
\def\hm{{\widehat m}}
\def\hn{{\widehat n}}
\def\hmu{{\widehat \mu}}
\def\hnu{{\widehat \nu}}

\def\p{\partial}
\def\D{\nabla}
\def\dk{\delta_\kappa}
\def\bu{\bullet}
\def\wt{\widetilde}
\def\wh{\widehat}
\def\ol{\overline}
\def\nn{\nonumber}
\def\<{\langle}
\def\>{\rangle}

\newcommand{\bk}[2]{\langle #1|#2 \rangle}

\begin{titlepage}

%%%%% Footnote symbol in title page %%%%%
\renewcommand{\thefootnote}{\fnsymbol{footnote}}

%%%%% report number %%%%%
\rightline{SU-ITP-08/36}

%%%%% Title %%%%%
\vspace{35mm} \baselineskip 9mm
\begin{center}
{\Large\bf Supersymmetric Heterotic Action out of M5 Brane}
\end{center}

%%%%% Authors and Addresses %%%%%
\baselineskip 6mm \vspace{10mm}
\begin{center}
{\large Jaemo Park}$^{1,2,3}$\footnote{\tt jaemo@postech.ac.kr} {and}
{\large Woojoo Sim}$^{1}$\footnote{\tt space@postech.ac.kr}
\\[5mm]

{\small\it
$^1$Department of Physics, POSTECH \\
Pohang 790-784, Korea \\ \vspace{0.3cm}

$^2$Postech Center for Theoretical Physics (PCTP), POSTECH \\
Pohang 790-784, Korea \\ \vspace{0.3cm}

$^3$Department of Physics, Stanford University \\
Stanford, CA 94305-4060, USA \\ \vspace{0.8cm}
}
\end{center}

%%%%% Abstract %%%%%
%\vfill
\vspace{20mm}
\begin{center}
{\bf Abstract}
\end{center}
\indent Generalizing the work by Cherkis and Schwarz~\cite{cherkis},
we carry out the double dimensional reduction of supersymmetric M5 brane on K3
to obtain the supersymmetric action of heterotic string in 7-dimensional flat space-time. 
Motivated by this result, we propose the supersymmetric heterotic action
in 10-dimensional flat space-time where the current algebra is realized in a novel way.
We explicitly verify the $\kappa$-symmetry of the proposed action.

\end{titlepage}

%%%%% Footnote symbol in body %%%%%
\renewcommand{\thefootnote}{\arabic{footnote}} \setcounter{footnote}{0}

\section{Introduction}

Understanding of BPS objects in string theory and M theory
shed great insight on the structure of string and M theory. 
The prominent example is D-brane in string theory, 
which facilitated great progress in understanding
the nonperturbative aspects of string theory. 
In M-theory, the counterpart to D-brane is M2 and M5-brane. 
Certainly the better understanding of these BPS objects would lead to
the enhanced understanding of M-theory but the status quo is
something to be desired compared with the understanding of D-brane.
In the current paper, we concentrate on the M5-brane
and make use of the M5-brane action, in particular, for one application.
One peculiar feature of M5-brane is that 
the worldvolume excitations contain self-dual 2-form fields.
In order to write a Dirac-Born-Infeld (DBI) type action, 
novel approaches are needed. 
In case the worldvolume is 6-d Minkowskian spacetime, 
one way is to give up the manifest Lorentz invariance and 
to write apparent non-covariant action with hidden Lorentz invariance~\cite{perry}. 
Another way is to impose the self-duality via gauge symmetry
with auxiliary fields~\cite{pasti}. 
In this formalism, after a suitable gauge fixing, 
the action is the same as the one proposed at ref.~\cite{perry}
as shown at ref.~\cite{park}. 
The supersymmetric action of the M5-brane in the flat background
was constructed in ref.~\cite{park}. 
M5-brane action in arbitrary curved background was constructed
in ref.~\cite{bandos} and its worldvolume field equations were developed
in ref.~\cite{bandos2,howe,sorokin}.

It is shown at ref.~\cite{cherkis} that M5-brane wrapping on K3
gives rise to the heterotic string in 7-dimension for the bosonic action. 
It is natural to consider the supersymmetric generalization of it. 
We explicitly construct the supersymmetric heterotic action in 10-d Minkowskian space
motivated by the action derived from the M5-brane wrapping on K3. 
There are various constructions of kappa symmetric heterotic
actions~\cite{tonin,tonin2,candiello,cederwall}. 
In each of the cases, the realization of the chiral current algebra is different
and it's interesting to show the equivalence of each approach.

The content of the paper is as follows. 
In section~\ref{sec:bosonicM5}, 
we review the bosonic M5-brane wrapped on K3~\cite{cherkis}. 
In section~\ref{sec:superM5}, we extend the reduction
to the case of the supersymmetric M5-brane action. 
The resulting action is the supersymmetric heterotic action
in 7-dimensional flat space. 
Interesting point is that the lattice derived from the two forms of K3
has the signature of $(19,3)$.
The zero modes of the self-dual two-form give rise to
22 scalars of the heterotic action. 
Among these scalars, only 3 scalars coming from $(0,3)$ part
have the supersymmetric extension while the other 19 scalars
have the same form as in the bosonic case. 
Although the dimensional reduction is carried
out at the leading orders of the fermionic variables, 
it is sufficient to give the clue to write the heterotic action in 10-d.
In section~\ref{sec:10d}, we write down
the supersymmetric heterotic action in 10-d flat space.
To realize the current algebra, we have to introduce 16 scalars. 
From the exercise of the section 3, 
we guess that the action of the scalar is the same as that of bosonic case. 
Indeed we can explicitly verify the kappa symmetry of the proposed heterotic action.
In the appendix we present the details of the proof of the kappa symmetry.

\medskip
\section{Reduction of the Bosonic M5-Brane:Review}
\label{sec:bosonicM5}

In this section, we review the bosonic M5-brane wrapped on K3, which
is introduced in ref.~\cite{cherkis}. The M5-brane action used here
is the one introduced in ref.~\cite{pasti} where the worldvolume
covariance is manifest. The action is given as~\footnote{We use the
following indices for the coordinates of the related manifolds. For
the M5-brane worldvolume and for the string worldsheet,
$\mu,\nu,\cdots$ and $\al,\be,\cdots$ are used, respectively. For
the 11-d and the 7-d target spaces, $M,N,\cdots$ and $m,n,\cdots$
are used, respectively. We use $i,j,\cdots$ for K3 surface. Finally,
hatted indices are used for the corresponding tangent spaces of the
manifolds.}
\begin{align}
\cL_1 &= -\sqrt{ -\det \biggl( G_{\mu\nu} +
     i\frac{\tH_{\mu\nu}}{\sqrt{-Gu^2}} \biggr)}, \\
\cL_2 &= -\frac{1}{4u^2} \tH^{\mu\nu}H_{\mu\nu\rho}u^\rho.
\end{align}
The worldvolume fields of the action are $X^M$, the coordinates of
the target 11-d spacetime, and $B_{\mu\nu}$, the chiral two-form
whose field strength $H_{\mu\nu\rho}$ is self-dual. In addition, to
make the general worldvolume covariance of the action manifest, an
auxiliary scalar field $a$ is introduced as $u_\mu = \p_\mu a$. In
the action, the bosonic background $e_M\!^\hM$ couples to the
worldvolume fields through the pullback
\begin{align}
G_{\mu\nu} =\p_\mu X^M e_M\!^\hM \p_\nu X^N e_N\!^\hN \eta_{\hM\hN}
    =\p_\mu X^M\p_\nu X^N g_{MN},
\end{align}
which is the induced worldvolume metric.\footnote{The worldvolume
indices in the action are raised and lowered by the induced
metric.} $\tH^{\mu\nu}$ in the action is defined as
\begin{align}
\tH^{\mu\nu} &= \frac{1}{6}\ep^{\mu\nu\rho\si\tau\la}
     H_{\rho\si\tau} u_\la.
\end{align}

The dimensional reduction here is a double dimensional one, which
means that we wrap the worldvolume on the compact part of the target
manifold (K3 here). Then, denoting the coordinates of K3 by $\si^i$,
we can take the static gauge $X^i=\si^i$ by using the worldvolume
covariance of the action, such that $\si^\mu=(\xi^\al,\si^i)$ and
$X^M=(X^m,\si^i)$ where $\xi^\al$ are the world sheet coordinates of
the resulting heterotic string action.

In reduction, we take the zero modes of the fields on K3. For the
chiral two-form $B_{\mu\nu}$, the zero modes are the harmonic forms
on K3. However, the harmonic zero-form $B_{\al\be}$ and the harmonic
one-form $B_{\al i}$ on K3 do not contribute to the reduction because
$\p_iB_{\al\be}=0$ and $B_{\al i}=0$. Then, the only contributing one
is the harmonic two-form on K3,
\begin{align} \label{B}
B_{ij}=\sum_{I=1}^{22} Y^I(\xi) b_{Iij}(\si),
\end{align}
where $B_{ij}$ is expanded into the 22 linearly independent
harmonic two-forms $b_{Iij}$ which form a basis of
$\rm{Harm}^2(\rm{K3},\mathbb Z) \cong H^2(\rm{K3},\mathbb Z)$.
Among $b_{Iij}$, three are self-dual and the rest are
anti-self-dual.

Then, by using eq.~\eqref{B} we find that the corresponding
nonvanishing components of $H_{\mu\nu\rho}$ and $\tH^{\mu\nu}$
are, respectively,
\begin{align} \label{H_bosonic}
H_{\al ij} = \sum_{I=1}^{22} \p_\al Y^I b_{Iij}, \quad
\tH^{ij} = \sum_{I=1}^{22} \sqrt{h} \tY^I (*b_I)^{ij},
\end{align}
where $h=\det h_{ij}$, $h_{ij}$ being the K3 metric, 
and $\tY^I = \ep^{\al\be}\p_\al Y^I u_\be$.

For the case of $X^M$ we can take $\p_i X^M=\de_i^M$ by using the
static gauge $X^i=\si^i$. This gives the decomposition of induced
metric as a block diagonal form $G_{\mu\nu}= \tG_{\al\be}\oplus
h_{ij}$, where $\tG_{\al\be}$ is the world-sheet induced metric. 
(As usual, the target manifold is a product one $M_7\times \rm{K3}$, 
$M_7$ being a 7-dimensional flat spacetime, so that
the metric becomes block diagonal: $g_{MN}= g_{mn}\oplus h_{ij}$.)

Then, inserting above nonvanishing components $H_{\al ij}$,
$\tG_{\al\be}$ and $h_{ij}$ into the action and integrating over
K3, the final form of the reduced action is given as
\begin{align} \label{bosonic}
S=-\sqrt{-\tG}\sqrt{1 + \frac{\tY^I M_{IJ}\tY^J}{\tG u^2}
    -\biggl(\frac{\tY^I L_{IJ}\tY^J}{2\tG u^2}\biggr)^2}
-\frac{\tY^I L_{IJ}\p_\al Y^J u^\al}{2u^2},
\end{align}
where
\begin{align} \label{LM}
L_{IJ} = \int_{K3} b_I\wedge b_J, \quad M_{IJ} = \int_{K3}
b_I\wedge *b_J.
\end{align}

Here the point is that the reduced action eq.~\eqref{bosonic} is the
dual of the heterotic string action compactified on $T^3$. The key
clue is given by the $Y^I b_{Iij}$ of eq.~\eqref{B} and by the
matrices $L$ and $M$ of eq.~\eqref{LM}. In fact, $b_I$ of K3 are the
basis of the self-dual lattice of signature $(19,3)$, which is
unique up to isometries, so that $L$ is the matrix of inner products
between the basis elements~\cite{serre,conway,aspinwall}. On the
string side, as the heterotic string winds around $T^3$, there
appear $16+3$ left-movers and 3 right-movers. These modes make the
even self-dual lattice of signature $(19,3)$, so called Narain
lattice. Thus, we can identify the Narain lattice with the lattice
arising from the M5-brane wrapped on K3, with the 22 movers being
$Y^I$. Moreover, $M$ can be interpreted as the matrix in which the
moduli of the torus are encoded. In fact, the 7-d action
eq.~\eqref{bosonic} can be directly shown to be equal to the
heterotic action compactified on $T^3$ as in ref.~\cite{cherkis}.
Note that $Y^I$ are the usual scalars in the above action
(\ref{bosonic}) but the additional gauge symmetry makes $Y^I$ to be
chiral.

\medskip
\section{Reduction of the Supersymmetric M5-brane}
\label{sec:superM5}

Now, let us wrap the supersymmetric M5-brane on K3, by extending the
bosonic case above, and see what the resulting 7-d theory is. Here
the fermionic coordinate $\th$ of the 11-d superspace and its
bilinears on K3 play the key role. We first review the
supersymmetric M5-brane action, and then perform the reduction. In
order to do it properly, we need the full supersymmetric M5-brane
action in an arbitrary supergravity background and carry out the
dimensional reduction. However we expect to obtain the heterotic
action in the flat space after the dimensional reduction and the
action should have the form
\begin{equation}
\cL = -\sqrt{-G}\sqrt{1 + \cdots}-\ep^{\al\be}
\bth\Ga_m\p_\al\th\p_\be X^m+\cdots.
\end{equation}
We can guess this from the Type I kappa symmetric action and bosonic
part of the heterotic action. The structure we do not know is how
the scalar degrees of freedom representing $E_8 \times E_8$ current
algebra are entered in the supersymmetric way. Thus our strategy is
to start from the supersymmetric M5-brane action in the flat space
and modify it in a minimal way so that it can describe M5-brane
action in an arbitrary supergravity background in leading order in
$\theta$. In fact, at the leading order of $\theta$, the fermionic
part coincides with the flat case as shown by de Wit for M2-brane~\cite{dewit}
and we expect the similar result for M5-brane.  Then we carry out
the dimensional reduction to see the structure of the current
algebra parts of the supersymmetric heterotic action in
7-dimensional flat space. With that information, it is easy to guess
the form of heterotic action in 10-dimensional flat space. We will
write that action in section~\ref{sec:10d} and explicitly verify the
kappa symmetry as well as supersymmetry.

\subsection{Supersymmetric M5-brane action}

To construct the supersymmetric M5-brane action, we need to extend
the target space to the superspace with the coordinates
$Z^A(\si)=(X^M,\th^a)$, where $\th(\si)$ is a 32-component Majorana
spinor, an irreducible representation in 11-d. Then, together with
the two-form field $B_{\mu\nu}$, both the bosonic and the fermionic
fields in the theory have the same correct on-shell degrees of
freedom, eight: five for $X^M$, three for $B_{\mu\nu}$; and eight
for $\th$, as expected for a brane of maximal supersymmetry. The
auxiliary scalar field $a(\si)$ keeps its role as making the general
covariance manifest. The concrete form of the action is
\cite{park,bandos}
\begin{align} \label{m5}
S_{M5}=\int d^6\si (\cL_1 + \cL_2) + S_{WZ},
\end{align}
where
\begin{align}
\cL_1 &= -\sqrt{ -\det \biggl( G_{\mu\nu}
            + i\frac{\tcH_{\mu\nu}}{\sqrt{-Gu^2}} \biggr)}, \nn \\
\cL_2 &= -\frac{1}{4u^2} \tcH^{\mu\nu}\cH_{\mu\nu\rho}G^{\rho\si}u_\si, \nn \\
S_{WZ} &=  \int \biggl( c_6 + \frac{1}{2}H\wedge c_3 \biggr).
\end{align}
Here, the worldvolume fields $Z^A(\si)$ and the 11-d supergravity
backgrounds enters the action via the pullbacks
\begin{align} \label{pb_gen}
\Pi_\mu^\hA &= \p_\mu Z^A E_A\!^\hA, \nn \\
c_{\mu\nu\rho} &= \p_\mu Z^C \p_\nu Z^B \p_\rho Z^A C_{ABC} , \nn \\
G_{\mu\nu} &= \Pi_\mu^\hM \Pi_\nu^\hN  \eta_{\hM\hN},
\end{align}
where $E_A\!^\hA$ and $C_{ABC}$ are the supervielbein and the
three-superform of the 11-d supergravity, respectively. In $S_{WZ}$,
$c_6$ is the pullback of $C_6$ whose field strength is the dual of
$dC_3$:
\begin{align}
*dC_3 = dC_6+\frac{1}{2}C_3\wedge dC_3.
\end{align}
The two-form field $B_{\mu\nu}$ enters the action via the
supersymmetrized field strength $\cH$:
\begin{align}
\cH = dB-c_3, \quad \tcH^{\mu\nu} =
\frac{1}{6}\ep^{\mu\nu\rho\si\tau\la}
     \cH_{\rho\si\tau} u_\la,
\end{align}
where $c_3$ is the pullback three-from in eq.~\eqref{pb_gen}.

Now, let us express the above pulled-back quantities by using the
supervielbein $E_A\!^\hA$ evaluated up to the 2nd order of
$\th$ and the three-superform $C_{ABC}$ taken up to the leading
order in $\theta$. Suppressing the gravitino and the 3-form gauge field
contributions, the components of $E_A\!^\hA$ and $C_{ABC}$ are given
as~\cite{dewit}
\begin{alignat}{2} \label{bg}
E_M\!^{\hM} &= e_M\!^\hM - \bth\Ga^\hM\om_M\th,
    &\quad E_M\!^\ha &= (\om_M\th)^\ha, \nn \\
E_a\!^\hM &= -(\bth\Ga^\hM)_a, & E_a\!^\ha &= \de_a\!^\ha, \nn \\
C_{MNP} &= 0, & C_{MNa} &= (\bth\Ga_{MN})_a,  \nn \\
C_{Mab} &= (\bth\Ga_{MN})_{(a}(\bth\Ga^N)_{b)},
    & C_{abc} &= (\bth\Ga_{MN})_{(a}(\bth\Ga^M)_b(\bth\Ga^N)_{c)},
\end{alignat}
where $e_M\!^\hM$ is the bosonic vielbein and
$\om_M=\frac{1}{4}\om_M^{\hP\hQ}\Ga_{\hP\hQ}$ is the spin
connection. Here we take essentially the leading order expression in
$\theta$ with minimal addition to make the partial derivative acting
on $\theta$ covariant derivative.

Then, by using eq.~\eqref{bg}, the components of the pullback
vielbein $\Pi_\mu^\hA$ in eq.~\eqref{pb_gen} are evaluated as
\begin{align} \label{pbv}
\Pi_\mu^\hM &= \p_\mu X^M E_M\!^\hM + \p_\mu\th^a E_a\!^\hM \nn \\
&= \p_\mu X^M e_M\!^\hM - \bth\Ga^\hM \nabla_\mu\th, \nn \\
\Pi_\mu^\ha &= \p_\mu X^M E_M\!^\ha + \p_\mu\th^a E_a\!^\ha \nn \\
&= \nabla_\mu\th^\ha,
\end{align}
where $\nabla_\mu\th = (\p_\mu+\om_\mu)\th$ and $\om_\mu=\p_\mu
X^M\om_M$, in which the worldvolume spin connection
$\om_\mu^{\hmu\hnu}$ (under static gauge) is encoded as well as the
$SO(5)$ gauge connection~\cite{kallosh}.

In a similar manner, the pullbacks $c_3$ and $c_6$ are evaluated
as
\begin{align} \label{pb_form}
c_3 &= \frac{1}{2} \bth \Ga_{MN} d\th
    \biggl( dX^M dX^N + \bth\Ga^M d\th dX^N
        + \frac{1}{3} \bth\Ga^M d\th \bth\Ga^N d\th \biggr), \nn \\
c_6 &= \bth\Ga_{M_1\cdots M_5}d\th
    \biggl(\frac{1}{5!}  dX^{M_1}\cdots dX^{M_5}
        + \frac{1}{48} \bth\Ga^{M_1}d\th  dX^{M_2}\cdots dX^{M_5} \biggr) \nn \\
    &\quad + \text{terms vanishing for the reduction on a 4-manifold},
\end{align}
where $dX^M = \p_\mu X^M d\si^\mu$ and $d\th = \D_\mu\th d\si^\mu$.
We emphasize again that there are higher order corrections in
$\theta$ for an arbitrary background but by concentrating on the
leading order corrections we get enough information to guess the
heterotic action in 10-dimensional flat space. As we will see later,
the resulting 7-dimensional string action obtained by leading order
approximation is likely to be exact, though we will not try to prove
its kappa symmetry.

The super M5-brane action in eq.~\eqref{m5} is invariant under the
kappa transformations of the worldvolume
fields~\cite{bandos,bandos2}
\begin{align}\label{kappa}
i_\ka \Pi^\ha = \de_\ka Z^A E_A\!^\ha = (1+\Ga)_\hb^\ha \ka^\hb,
\quad i_\ka \Pi^\hM = 0, \quad \dk \cH = -i_\ka dc_3, \quad \dk a =
0,
\end{align}
where $i_\ka$ denotes the pullback of the interior product on a
background superform with respect to $\dk Z^A$. In
eq.~\eqref{kappa}, $\Ga$ is determined to satisfy $\Ga^2=1$ for the
kappa invariance of the action:
\begin{align}
\Ga = \frac{1}{\cL_1}\biggl(
    \bga + \frac{1}{2u^2} \tcH^{\mu\nu}\ga_{\mu\nu}\ga^\rho u_\rho
    + \frac{1}{16u^2}\ep_{\mu\nu\rho\la\tau\si}\tcH^{\mu\nu}\tcH^{\rho\la}\ga^{\si\tau}
    \biggr),
\end{align}
with $\ga_\mu=\Ga_\hM \Pi_\mu^\hM$ and $\bga=\ga_{012345}$. In the
flat background, the kappa variations above reduce to
\begin{align}
\de_\ka \th = (1+\Ga)\ka,
\quad \de_\ka X^M = -\de\bth\Ga^M\th,
\quad \de_\ka \cH_{\mu\nu\rho} = 6\de\bth\ga_{[\mu\nu}\p_{\rho]}\th.
\end{align}

\subsection{Decomposition of a 11-d spinor}

Before we perform the reduction of the supersymmetric action, let
us first examine the decomposition of 11-d spinor $\th$ into 7-d
and 4-d ones. The decomposition of {\bf 32} Majorana
representation of $Spin(10,1)$ under $Spin(6,1)\times Spin(4)$ is
\begin{align} \label{dec_32}
{\bf 32} = ({\bf 8}, {\bf 2}) + ({\bf 8}, {\bf 2'}),
\end{align}
where ${\bf 2}$ and ${\bf 2'}$ are respectively positive and negative
chirality Weyl spinors in 4-d, each of which is self-conjugate. As
will be described in section~\ref{sec:reduction}, we take the 4-d
spinors to be covariantly constant in the dimensional reduction. Then,
since the holonomy group of K3 is $SU(2)$, which is a subgroup of
$Spin(4)= SU(2)\times SU(2)$, 
only one of ${\bf 2}$ and ${\bf 2'}$ is covariantly constant 
according to whether K3 is self-dual or anti-self-dual. 
Therefore, taking ${\bf 2}$ to be covariantly constant, 
only $({\bf 8}, {\bf 2})$ of ${\bf 32}$ contributes to the
reduction on K3, and in 7-d there are 16 real spinor degrees of
freedom.

Now, let us express $({\bf 8}, {\bf 2})$ in terms of irreducible
representations of $Spin(6,1)$ and $Spin(4)$. Note first that the
spinors of $Spin(6,1)$ and $Spin(4)$ cannot be the Majorana, but can
be the symplectic Majorana (SM)~\cite{kugo,kleppe}. The SM
representation is given as a $USp(2)$ doublet, which is a pair of
spinors $\psi_1$ and $\psi_2$ that satisfy
\begin{align} \label{sm}
(\psi_A)^*=B\ep^{AB}\psi_B, \quad B^*B=-1,
\end{align}
with $\ep^{12}=-\ep^{21}=1$. (In the Majorana representation,
$\psi^*=B\psi$ with $B^*B=1$.) Here, $B$ is a unitary matrix that
relates $\Ga_\mu$ to $\Ga^*_\mu$, both of them being the
representations of the Clifford algebra, as
\begin{align}
\Ga^*_\mu = \et B\Ga_\mu B^{-1}, \quad \et=\pm 1,
\end{align}
where the sign of $\et$ depends on the signature of the metric.

In the case of $Spin(4)$, where the Weyl condition is available, the
two spinors in eq. \eqref{sm} can be the Weyl spinors of the same
chirality. Therefore, the irreducible spinor representation of
$Spin(4)$ is symplectic Majorana-Weyl (SMW). 

Now, considering above irreducible spinor representations in 7-d and
in 4-d, we can express the 11-d spinor $\th$, which is $({\bf 8},
{\bf 2})$ of ${\bf 32}$, as
\begin{align} \label{th_dec}
\th= \la_1\otimes\ch_1 + \la_2\otimes\ch_2,
\end{align}
where $\la_A$ and $\ch_A$ denote $Spin(6,1)$ and $Spin(4)$ spinors
satisfying the SM and the SMW (of positive chirality) conditions, respectively:
\begin{align} \label{sm_dec}
(\la_A)^*=B_7\ep^{AB}\la_B, \quad (\ch_A)^*=B_4\ep^{AB}\ch_B,
\end{align}
where $B_7^*B_7 = -1$ and $B_4^*B_4 = -1$.

The SM representation as an $USp(2)=SU(2)$ doublet implies that the
7-d action has $SU(2)$ symmetry in the sense $SU(2)=SO(4)/SU(2)$
where $SO(4)$ and $SU(2)$ in r.h.s. are, respectively, the Lorentz
and the holonomy group of K3. Then we need to assign the $SU(2)$
indices properly to $\la$ and $\ch$ according to their
transformation properties under $SU(2)$: 
lower indices to the ones transforming as fundamentals 
and upper indices to the ones transforming as complex conjugates. 
The result is
\begin{align}\label{index}
\bla^A = \ol{\la_A} \quad (\la^{*A}=(\la_A)^*), \quad \bch^A =
(\chi_A)^\dag \quad (\ch^{*A}=(\ch_A)^*).
\end{align}
This yields for 11-d spinor $\th$ and its Dirac conjugate $\bth$
\begin{align} \label{th_conj}
\th &= \la_1\otimes\ch_1 + \la_2\otimes\ch_2, \nn \\
\bth &= ((\la_1)^\dag\otimes(\ch_1)^\dag +
    (\la_2)^\dag\otimes(\ch_2)^\dag) \tGa_0\otimes\tGa_{(5)} \nn \\
&= \ol{\la_1}\otimes(\ch_1)^\dag +
    \ol{\la_2}\otimes(\ch_2)^\dag \nn \\
&= \bla^1\otimes\bch^1 + \bla^2\otimes\bch^2,
\end{align}
so that the $SU(2)$ invariant bilinear $\bth\th$ is expressed as
\begin{align} \label{scalar}
\bth\th = \sum_{A,B}\bla^A\la_B\bch^A\ch_B = \bla^A\la_A.
\end{align}
Here we have used the decompositions of 11-d gamma matrices
\begin{align} \label{gamma_dec}
\Ga_m = \tGa_m \otimes \tGa_{(5)}, \quad \Ga_i = 1 \otimes \tGa_i,
\end{align}
where $\tGa_m$ are the 7-d gamma matrices while $\tGa_i$ the 4-d
ones with $\tGa_{(5)}=-\tGa_{78910}$.

To show the above expression of $\th$ in eq.~\eqref{th_dec} is a consistent one,
we should check $\th$ can be a Majorana spinor with the expression. 
The complex conjugation of $\th$ becomes
\begin{align}
\th^* &= \sum_A \la^{A*}\otimes\ch^{A*} = \sum_A
B_7\ep^{AB}\la_B\otimes B_4\ep^{AC}\ch_C = (B_7\otimes B_4)\th.
\end{align}
Then we can expect $B_{11}=B_7\otimes B_4$ to be the conjugation
matrix in 11-d, satisfying the Majorana condition. In fact,
\begin{align}
B_{11}^*B_{11} &= B_7^*B_7\otimes B_4^*B_4 = -1\otimes -1 = 1, \nn \\
B_{11}^\dag B_{11} &= B_7^\dag B_7\otimes B_4^\dag B_4 = 1,
\end{align}
which shows that the 11-d Majorana condition, along with the 7-d SM
and the 4-d SMW conditions, is satisfied.

In the tangent space bases of K3 and $M_7$, the gamma matrices and
the corresponding conjugation matrices can be chosen as follows.
In the case of K3, taking its coordinate to be
$X^7,\cdots,X^{10}$,
\begin{align}
\tGa_{7,8,9} = \si_{1,2,3}\otimes \si_2, \quad \tGa_{10} =
1\otimes\si_1,
\end{align}
and for $M_7$,
\begin{align}
&\tGa_{0,1,2} = \si_{1,2,3}\otimes\tGa_{(5)}, \quad \tGa_{3,4,5,6}
= 1\otimes\tGa_{7,8,9,10},
\end{align}
where $\tGa_0$ has an extra factor $i$. The conjugation matrices
$B_4$ and $B_7$ that are compatible with the above gamma matrices
can be
\begin{align}
B_4 = i\si_2\otimes 1, \quad B_7 = \si_3\otimes\si_2\otimes 1.
\end{align}

The covariantly constant spinors on K3 can be understood
alternatively in the context of spin states that are created and
annihilated by the gamma matrices associated with the Hermitian
metric~\cite{green}. With the complex dimension being two, there are
two creation (annihilation) operators $\tGa_\ba$ ($\tGa_a$). Then,
the two covariantly constant spinors are identified as the lowest
state $|\Om\>$ and 
the highest state $|\bOm\>\sim\tGa_{12}|\Om\>$, respectively. 
This is because these two spin states interact with the $U(1)$ part 
of the spin connection, which is trivial on K3 (and on any Calabi-Yau $n$-fold).

On K3, the two spin states $|\Om\>$ and $|\bOm\>$ are of the same
positive chirality, because the raising operator $\Ga_\ba$ flips
the chirality and there are two such flips from $|\Om\>$ to
$|\bOm\>$. Besides, since $\Ga_a|\Om\>=\Ga_\ba|\bOm\>=0$, we can
take the two SMW spinors $\ch_1$ and $\ch_2$ in eq.~\eqref{sm_dec}
to be $|\Om\>$ and $|\bOm\>$, respectively. This property is used
in the following section to identify the spinor bilinears that
appear in the reduction.

\subsection{Reduction of the supersymmetric action}
\label{sec:reduction}

Now let us perform the double dimensional reduction of the
supersymmetric M5-brane action in eq.~\eqref{m5} on K3. Before
integrating over K3, let us first take the zero modes of the fields
and find the components that contribute to the action. For the
bosonic fields $B_{\mu\nu}$ and $X^M$ we can take their zero modes
as in the bosonic case. For the fermionic field $\th$, we can take
its zero mode to satisfy
\begin{align} \label{zm_th}
\D_i\th = (\p_i+\tfrac{1}{4}\om_i^{\hi\hj}\Ga_{\hi\hj})\th = 0,
\end{align}
i.e., covariantly constant on K3, since the spin connection
$\om_i^{\hM\hN}$ in the equation of motion reduces to
$\om_i^{\hi\hj}$ in eq.~\eqref{zm_th} with the condition
$\om_i^{\hm\hn}=\om_i^{\hj\hn}=0$ given by the product manifold
ansatz $e_i\!^\hm=\p_\al e_i\!^\hi=0$.~\footnote{In reduction, we
take $g_{MN}(X^M)=g_{mn}(X^m)\oplus g_{ij}(X^i)$ so that
$g_{mi}=\p_m g_{ij}=\p_i g_{mn}=0$. That is, $e_\al\!^\hi =e_i\!^\hm
=\p_\al e_i\!^\hi =\p_i e_m\!^\hm =0$ for vielbeins.}

By using $\D_i\th=0$ as well as the bosonic zero modes,
we can evaluate the components of the pulled-back quantities in the action.

First, the components of pullback vielbein $\Pi_\mu^\hM$ in eq.~\eqref{pbv} become
\begin{alignat}{2} \label{pbv_red}
\Pi_\al^\hm &= \p_\al X^m e_m\!^\hm - \bth\Ga^\hm\D_\al\th,
&\quad \Pi_\al^\hi &= -\bth\Ga^\hi\D_\al\th, \nn \\
\quad \Pi_i^\hm &= e_i\!^\hm =0, &\quad \Pi_i^\hi &= e_i\!^\hi.
\end{alignat}
Here, for $e_\al\!^\hi=\p_i e_m\!^\hm=0$,
$\D_\al=\p_\al+\tfrac{1}{4}\om_\al^{\hm\hn}\Ga_{\hm\hn}$,
$Spin(6,1)$ spin connection being pulled-back to 2-d worldsheet.
However, since we take $M_7$ to be Minkowskian, 
the connection is trivial so that $\D_\al=\p_\al$.

Then, the components of $G_{\mu\nu}$ are given by using
eq.~\eqref{pbv_red} as
\begin{align} \label{G_old}
G_{\al\be} = \tG_{\al\be} + \bth\Ga_i\p_\al\th \bth\Ga^i\p_\be\th,
\quad G_{ij} = h_{ij}, \quad G_{\al i} = -\bth\Ga_i\p_\al\th,
\end{align}
where $\tG_{\al\be} =\eta_{\hm\hn}\Pi_\al^\hm\Pi_\be^\hn$ is the
induced metric of the 2-d worldsheet in the flat 7-d superspace background, 
and $h_{ij}$ is the metric of K3 surface.

Next, we get the components of $c_3$ in eq.~\eqref{pb_form} as
\begin{align}
c_{\al ij} = \frac{1}{2}\bth\Ga_{MN}\p_\al\th\de_{[i}^M\de_{j]}^N
    = \bth\Ga_{ij}\p_\al\th,
\quad c_{\al\be\ga} = c_{ijk} = 0.
\end{align}
Although there is another nonzero term $c_{\al\be i}$, it
does not contribute to the action. This can be anticipated from
the observation that $\tc^{ij}$ are the only nonvanishing
components of $\tc^{\mu\nu} =\frac{1}{6}
\ep^{\mu\nu\rho\si\tau\la} c_{\rho\si\tau} u_\la$ as well as
$H^{ij}$ in $\cL_2$, and that $H_{\al ij}\wedge c_{\al\be i}=0$ in
$S_{WZ}$. That is, the only nonvanishing component of
$\cH_{\mu\nu\rho}$ and $\tcH^{\mu\nu}$ are, respectively,
\begin{align}
\cH_{\al ij} = H_{\al ij} - c_{\al ij}, \quad \tcH^{ij} = \tH^{ij}
- \tc^{ij}.
\end{align}

Finally, $c_6$ is evaluated in the similar manner as
\begin{align} \label{c6}
c_6
= -\ep^{\al\be} \bigg( \bth\Ga_{(5)}\Ga_m\p_{\al}\th \p_\be X^m
    + \frac{1}{2} \bth\Ga_{(5)}\Ga_m\p_{\al}\th \bth\Ga^m\p_\be\th\bigg),
\end{align}
where $\Ga_{(5)} = -\Ga_{78910}$.

Now, note that the nonvanishing components of the above pullbacks
$G_{\mu\nu}$, $c_3$, and $c_6$, include three kinds of $\th$
bilinears:
\begin{align}
\bth\Ga_m\p_\al\th \;(\text{in $c_6$}), \quad \bth\Ga_i\p_\al\th
\;(\text{in $G_{\mu\nu}$}), \quad \bth\Ga_{ij}\p_\al\th \; (=c_3).
\end{align}
($\bth\Ga_{(5)}\Ga_m\p_{\al}\th$ in $c_6$ is identified to be the same
as $\bth\Ga_m\p_{\al}\th$ as shown below.) 
To complete the reduction of the pullbacks, we need to identify these bilinears
using the decomposition of $\th$ in eq.~\eqref{th_conj}
and the properties of the covariantly constant SMW spinors $\ch_A$ on K3. 
The gamma matrix decompositions introduced in eq.~\eqref{gamma_dec} are used here.

First, let us evaluate $\bth\Ga_m\p_\al\th$ in $c_6$:
\begin{align} \label{bi0_old}
\bth\Ga_m\p_\al\th &=\sum_{A,B}(\bla^A\otimes\bch^A)
    (\tGa_m\otimes\tGa_{(5)})(\p_\al\la_B\otimes\ch_B) \nn \\
&=\sum_{A,B}\bla^A\tGa_m\p_\al\la_B \bch^A\ch_B,
\end{align}
where we have used $\bch^1\tGa_{(5)}=\bch^1$ and $\bch^2\tGa_{(5)}=\bch^2$,
$\ch_1$ and $\ch_2$ being positive chirality spinors. 
Note that in the last line of eq.~\eqref{bi0_old} appear
the bilinear scalars $\bch^A\ch_B$ on K3. 
Reminding that $\ch_1$ and $\ch_2$ are respectively
the lowest and the highest (2nd excited) spin states and $\bch^A=(\ch_A)^\dag$,
we can identify those bilinears on K3: 
$\bch^1\ch_2=\bch^2\ch_1=0$ by the orthogonality between the states;
and $\bch^1\ch_1=\bch^2\ch_2=1$ by the constancy of $\chi_A$. 
Then, eq.~\eqref{bi0_old} becomes
\begin{align} \label{bi0}
\bth\Ga_m\p_\al\th = \bla^A\tGa_m\p_\al\la_A.
\end{align}
By using this result we can complete the reduction of $c_6$ which is
the first term of $S_{WZ}$. However, note that
$\bth\Ga_{(5)}\Ga_m\p_{\al}\th=\bth\Ga_m\p_{\al}\th$ as can be seen in
eq.~\eqref{bi0_old}, $\ch_A$ being of positive chirality, so that in
eq.~\eqref{c6}
\begin{align}
\ep^{\al\be}\bth\Ga_{(5)}\Ga_m\p_{\al}\th \bth\Ga^m\p_\be\th =
\ep^{\al\be}\bth\Ga_m\p_{\al}\th \bth\Ga^m\p_\be\th = 0.
\end{align}
As a result, inserting eq.~\eqref{bi0} into eq.~\eqref{c6} and
integrating over K3,
\begin{align} \label{L3}
\int c_6 = -\ep^{\al\be}
    (\bla^A\tGa_m\p_\al\la_A) \p_\be X^m.
\end{align}

Next, let us evaluate $\bth\Ga_i\p_\al\th$ which appears in
$G_{\al\be}$ and in $G_{\al i}$:
\begin{align} \label{bi1_old}
\bth\Ga_i\p_\al\th &=\sum_{A,B}(\bla^A\otimes\bch^A)
    (1\otimes\tGa_i)(\p_\al\la_B\otimes\ch_B) \nn \\
&=\sum_{A,B}\bla^A\p_\al\la_B \bch^A\tGa_i\ch_B.
\end{align}
Here, the 4-d bilinears $\bch^A\tGa_i\ch_B$ in the last line are the
one-forms on K3. Similarly to the above bilinear scalars, we can
identify these bilinears by using the spin sate analysis. 
But here we should consider the gamma matrices in the bilinears, 
which are the ladder operators acting on the spin states.
For example, $\bch^1\tGa_\ba\ch_1=0$, because $\tGa_\ba\ch_1$ is the first
excited state which is orthogonal to the lowest state $\ch_1$.
Similarly, $\bch^1\tGa_a\ch_2=0$ where $\tGa_a\ch_2$ is also the
first excited state~\footnote{To avoid confusion, 
we denote the complex coordinates of K3 by $\om^a$ and $\om^\ba$, 
while the real coordinates are denoted by $\si^i$.}. 
Note, besides orthogonality, $\Ga_a\ch_1=\Ga_\ba\ch_2=0$ gives vanishing bilinears.
Then, examining the bilinear one-forms above in this way,
we find that all of them vanish, which means,
\begin{align} \label{bi1}
\bth\Ga_i\p_\al\th =0.
\end{align}
This makes $G_{\al\be}$ and $G_{\al i}$ of eq.~\eqref{G_old}
become
\begin{align}
G_{\al\be} = \tG_{\al\be}, \quad G_{\al i} =0.
\end{align}
That is, $G_{\mu\nu}$ becomes block diagonal as in the bosonic
case:
\begin{align} \label{G}
G_{\mu\nu} = \tG_{\al\be} \oplus h_{ij}, \quad G=\tG h,
\end{align}
which implies that we get the same reduced expressions
as in the bosonic case for $G$ and for the terms contracted by $G_{\mu\nu}$.

Finally, let us evaluate $c_{\al ij}= \bth\Ga_{ij}\p_\al\th$, which
will be verified to play a key role in the reduction. It becomes
\begin{align} \label{bi2_old}
\bth\Ga_{ij}\p_\al\th &=\sum_{A,B}(\bla^A\otimes\bch^A)
    (1\otimes\tGa_{ij})(\p_\al\la_A\otimes\ch_A) \nn \\
&=\sum_{A,B}\bla^A\p_\al\la_B \bch^A\tGa_{ij}\ch_B.
\end{align}
In this case, the bilinears $\bch^A\tGa_{ij}\ch_B$ are the two-forms on K3.
As in the previous cases, we can easily check 
which (components) of these two-forms vanish,
by considering the successive ladder operators $\tGa_a$ ($\tGa_{\ba})$
between the spin states. For example,
$\bch^1\tGa_{\ba\bb}\ch_1$, the antiholomorphic components of
$\bch^1\tGa_{ij}\ch_1$, vanish:
\begin{align}
\bch^1\tGa_{\ba\bb}\ch_1
    \sim \bch^1(\tGa_\ba\tGa_\bb-\tGa_\bb\tGa_\ba)\ch_1
    \sim \bch^1\ch_2 = 0,
\end{align}
where the lowest state $\ch_1$ has been raised by the two
successive raising operators $\tGa_\ba$ and $\tGa_\bb$
to be proportional to the highest state $\ch_2$. 
In this way, we can find the vanishing components of each bilinear two-form
and can write down the two-forms with the nonvanishing components only.
The result is
\begin{align} \label{bi2_nonzero}
\bch^1\tGa_{ij}\ch_1 d\si^i\wedge d\si^j
    &= \bch^1\tGa_{a\bb}\ch_1 d\om^a\wedge d\om^{\bb}, \nn \\
\bch^2\tGa_{ij}\ch_2 d\si^i\wedge d\si^j
    &= \bch^2\tGa_{a\bb}\ch_2 d\om^a\wedge d\om^{\bb},  \nn \\
\bch^1\tGa_{ij}\ch_2 d\si^i\wedge d\si^j
    &= \bch^1\tGa_{ab}\ch_2 d\om^a\wedge d\om^b, \nn \\
\bch^2\tGa_{ij}\ch_1 d\si^i\wedge d\si^j
    &= \bch^2\tGa_{\ba\bb}\ch_1 d\om^{\ba}\wedge d\om^{\bb}.
\end{align}

Now, let us examine what these two-forms, which constitute $c_{\al
ij}$, are on K3 and which role they play in the reduction of
$\cH_{\al ij} = H_{\al ij}-c_{\al ij}$ and of $H_{\al ij}\wedge
c_{\al ij}$.

First, we can show that $\bch^1\tGa_{a\bb}\ch_1 d\om^a\wedge d\om^{\bb}$ and $\bch^2\tGa_{a\bb}\ch_2  d\om^a\wedge d\om^{\bb}$ in eq.~\eqref{bi2_nonzero}
are proportional to the K\"{a}hler form $J=ig_{a\bb}d\om^a\wedge d\om^\bb$.
This is realized as
\begin{align}
\bch^1\tGa_{a\bb}\ch_1
    &= \bch^1 g_{a\bb}\ch_1 - \bch^1\tGa_\bb\tGa_a\ch_1 = g_{a\bb}, \nn \\
\bch^2\tGa_{a\bb}\ch_2
    &= -\bch^2 g_{a\bb}\ch_2 + \bch^2\tGa_a\tGa_\bb\ch_2 = -g_{a\bb},
\end{align}
where we have used $\tGa_a\ch_1=\tGa_\bb\ch_2=0$ and normalization
$\bch^1\ch_1=\bch^2\ch_2=1$, $\ch_1$ and $\ch_2$ being covariantly
constant. One property of the K\"{a}hler form $J$ is that it is
self-dual and harmonic. It is self-dual because, up to a positive
constant factor,
\begin{align}
*g_{a\bb}d\om^a\wedge d\om^\bb \sim -\sqrt{g} g_{a\bb}
\ep^a_\bc\ep^\bb_d d\om^\bc\wedge d\om^d \sim g_{d\bc} d\om^d\wedge
d\om^\bc
\end{align} %%% Check the sign!!!
where the Hermitian property of the metric, $g_{ab}=g_{\ba\bb}=0$
has been used. Then since K\"{a}hler form $J$ is closed and is
self-dual as shown in above, it is coclosed as well. Therefore, it
is harmonic.

One can show $\Om=\bch^1\tGa_{ab}\ch_2 d\om^a\wedge d\om^b$ and
$\bOm=\bch^2\tGa_{\ba\bb}\ch_1 d\om^{\ba}\wedge d\om^{\bb}$ 
in eq.~\eqref{bi2_nonzero} are also self-dual and harmonic by a similar manner. 
Thus we have identified the nonvanishing bilinears in eq.~\eqref{bi2_nonzero}
as three independent self-dual harmonic two-forms $J$, $\Om$ and $\bOm$ on K3. 
These three self-dual two-forms can be transformed into the real ones,
$J$, $\frac{1}{2}(\Om+\bOm)$ and $\frac{1}{2i}(\Om-\bOm)$, 
which are the self-dual elements of $H^2(\rm{K3},\mathbb{R})$. 

Then, the point is that we can take the harmonic two-form basis $b_{Iij}$ on K3,
consisting of 19 anti-self-dual forms and 3 self-dual forms, 
such that the above three real harmonic two-forms are
the self-dual elements of the basis\footnote{We denote the anti-self-dual
and the self-dual harmonic two-forms by $b^-_I$ and $b^+_I$, respectively.}:
\begin{align} \label{sd_basis}
b^+_1 = J, \quad
b^+_2 = \frac{1}{2}(\Om+\bOm), \quad 
b^+_3 = \frac{1}{2i}(\Om-\bOm).
\end{align}

As a result of the above identification of the bilinear
two-forms in eq.~\eqref{bi2_nonzero}, we can rewrite $c_{\al
ij}=\bth\Ga_{ij}\p_\al\th$ of eq.~\eqref{bi2_old} as
\begin{align} \label{bi2}
c_{\al ij} = \sum_{I=1}^{3} c_\al^I b^+_{Iij}
\end{align}
where we have rearranged the 7-d bilinears $\bla^A\p_\al\la_B$ in
eq.~\eqref{bi2_old} to be the coefficients $c_\al^I$ of the
self-dual basis $b^+_{I}$:
\begin{align} \label{c}
c_\al^{1} &= -i(\bla^1\p_\al\la_1 - \bla^2\p_\al\la_2), \nn \\
c_\al^{2} &= \bla^1\p_\al\la_2 + \bla^2\p_\al\la_1, \nn \\
c_\al^{3} &= i(\bla^1\p_\al\la_2 - \bla^2\p_\al\la_1).
\end{align}
That is, $c_{\al ij}$ is expanded into only the self-dual part of
the harmonic basis $H^2(\rm{K3},\mathbb{R})$, 
whereas $H_{\al ij}$ is expanded into the whole basis as in eq.~\eqref{H_bosonic}.
As a result, $\cH_{\al ij}=H_{\al ij} - c_{\al ij}$, 
the only contributing components of $\cH_{\mu\nu\rho}$, can be expanded as
\begin{align} \label{H}
\cH_{\al ij} = \sum_{I=1}^{19} \p_\al Y^I b^-_{Iij}
    + \sum_{I=1}^{3} Z_\al^I b^+_{Iij},
\end{align}
where the coefficients of the self-dual part are the supersymmetric
quantities,
\begin{align}
Z_\al^I = \p_\al Y^I - c_\al^I.
\end{align}
Then, $\tcH^{ij}$, the only nonvanishing components of
$\tcH^{\mu\nu}$ become
\begin{align} \label{Ht}
\tcH^{ij} = \sqrt h
    \bigg(\sum_{I=1}^{19} \tY^I (*b^-_I)^{ij}
        + \sum_{I=1}^{3} \tZ^I (*b^+_I)^{ij} \bigg),
\end{align}
where $\tZ^I = \ep^{\al\be}Z_\al^I u_\be$.

We can interpret this result in terms of the dual description of
the heterotic string as in the bosonic case. In fact, the
self-duality of $H_{\al ij}$ forces it to be expanded as
\begin{align} \label{H_sd}
H_{\al ij} = \sum_{I=1}^{19} \p_\al Y_-^I b^-_{Iij}
    + \sum_{I=1}^{3} \p_\al Y_+^I b^+_{Iij}
\end{align}
where $Y_-^I$ and $Y_+^I$ corresponds to the left-movers and the
right-movers of the heterotic string, respectively. (We will use
terms left movers and right movers for  $Y_-^I$ and $Y_+^I$ but this
terminology is more appropriate only after the gauge fixing. The
situation is the same for the self-duality of $H_{\al ij}$, which is
manifest after gauge fixing.) Then, when the heterotic string is
compactified on $T^3$, the left movers $Y_-^I$ which is bosonic in
10-d should remain bosonic, whereas the right-movers $Y_+^I$ should
remain supersymmetric as in 10-d. This is well satisfied by
eq.~\eqref{H}, where $c_\al^I$, the fermionic bilinears in 7-d, are
added only to the three right-movers $Y_+^I$ making $Z_\al^I$
supersymmetric.

Now if we insert $\cH_{\al ij}$ and $\tcH^{ij}$ into $\cL_1$ and
$\cL_2$ of the supersymmetric action, as in the bosonic case, the
related terms are written in terms of $\int b_I\wedge b_J$ and
$\int b_I\wedge *b_J$. However, note
\begin{align}
\int b^-_I\wedge b^+_J = \int b^-_I\wedge *b^+_J
= \int b^+_J\wedge *b^-_I =\int  -b^-_I\wedge b^+_J = 0.
\end{align}
That is, the product of a anti-self-dual two-form and a self-dual
two-form vanishes. As a result, as $\cL_1$ and $\cL_2$ are
integrated over K3, $L_{IJ}=\int_{K3} b_I\wedge b_J$ and
$M_{IJ}=\int_{K3} b_I\wedge *b_J$ of eq.~\eqref{LM} become block
diagonal:
\begin{align} \label{LM_block}
L = L^- \oplus L^+, \quad M = M^- \oplus M^+,
\end{align}
where $L^-$ ($M^-$) and $L^+$ ($M^+$) are the blocks constructed from
the anti-self-dual and the self-dual elements of the harmonic basis,
respectively: $L^-_{IJ} = \int_{K3} b^-_I\wedge b^-_J$, $L^+_{IJ} =
\int_{K3} b^+_I\wedge b^+_J$; and similarly for $M^-_{IJ}$,
$M^+_{IJ}$.

The above decomposition of $L$ and $M$ makes all terms including $L$
and $M$ decomposed into the terms including the coefficients of the
self-dual forms and those of the anti-self-dual forms. This implies
that the left-movers and the right-movers in eq.~\eqref{H} are
decoupled in the action as expected for the dual heterotic string
action. For example, $\cL_2$ becomes
\begin{align}
\int_{K3} \tcH^{ij}\cH_{\al ij}u^\al
    &\sim \int_{K3} \sqrt h (\tY^I *(*b^-_I\wedge *b^-_J)\p_\al Y^J
        + \tZ^I *(*b^+_I\wedge *b^+_J)Z_\al^J) \nn \\
    &\sim (\tY^I L^-_{IJ} \p_\al Y^J + \tZ^I L^+_{IJ} Z_\al^J)u^\al,
\end{align}
where we can see the terms of left-movers, $\tY^I L^-_{IJ} \p_\al
Y^J$ and those of right-movers, $\tZ^I L^+_{IJ} Z_\al^J$ are
separated.

In this way, inserting $G_{\mu\nu}$ of eq.~\eqref{G} besides
$\cH_{\al ij}$ and $\tcH^{ij}$ into $\cL_1$ and $\cL_2$, and
integrating over K3,
\begin{align} \label{L1L2}
\cL_1 &\rightarrow -\sqrt{-G}
    \sqrt{1 + \frac{\tY M^-\tY + \tZ M^+\tZ}{Gu^2}
        + \biggl(\frac{\tY L^-\tY + \tZ L^+\tZ}{2Gu^2}\biggr)^2}, \nn \\
\cL_2 &\rightarrow -\frac{(\tY L^- \p_\al Y + \tZ L^+ Z_\al)
u^\al}{2u^2},
\end{align}
where we have omitted the basis indices $I,J$.

Thus we have completed the reduction of $\cL_1$ and $\cL_2$. Besides,
we have already reduced the first term of $S_{WZ}$ in eq.~\eqref{L3}.
Then the rest part is the second term of $S_{WZ}$, which is
$\frac{1}{2}H\wedge c_3$. This term is easily reduced by using
$c_{\al ij}$ in eq.~\eqref{bi2} and $H_{\al ij}$ in eq.~\eqref{H_bosonic}:
\begin{align} \label{L4}
\frac{1}{2} \int_{K3} H\wedge c_3
&= \frac{1}{4}  \int_{K3} \ep^{\al\be}\ep^{ijkl}H_{\al ij}c_{\be ij} \nn \\
&= \int_{K3} \ep^{\al\be} \sqrt h \p_\al Y^I (b^+_I\wedge b^+_J) c_\be^J \nn \\
&= \ep^{\al\be} \p_\al Y L^+ c_\be.
\end{align}
Note that this term has only the block of the self-dual basis.

Finally, collecting the results of the reductions, eq.~\eqref{L3},
\eqref{L1L2}, and \eqref{L4}, the full action of the 7-d heterotic
string becomes
\begin{align} \label{7d_action}
\cL_1 &= -\sqrt{-G}
    \sqrt{1 + \frac{\tY M^-\tY + \tZ M^+\tZ}{Gu^2}
        + \biggl(\frac{\tY L^-\tY + \tZ L^+\tZ}{2Gu^2}\biggr)^2}, \nn \\
\cL_2 &= -\frac{(\tY L^- \p_\al Y + \tZ L^+ Z_\al) u^\al}{2u^2}, \nn \\
\cL_3 &= -\ep^{\al\be}
    (\bla^A\tGa_m\p_\al\la_A)\p_\be X^m, \nn \\
\cL_4 &= -\ep^{\al\be} (\p_\al Y L^+ c_\be).
\end{align}

\medskip
\section{10-d Heterotic String Action and Its Kappa Symmetry}
\label{sec:10d}

We now try to construct the 10-d action of the heterotic string by
using the structure of the 7-d action eq.~\eqref{7d_action}. The
10-d action can be naturally constructed by considering the
intersection matrix $L$ above along with the left-moving and the
right-moving part of the 7-d action.

As mentioned in the bosonic reduction, the topology of $K3$ requires
$L=\int_{K3}b_I\wedge b_J$ to be the intersection matrix of the even
self-dual lattice with signature $(19,3)$. This lattice is unique up
to isometries which preserve the inner product, so that one can
choose $L$ as~\cite{aspinwall}
\begin{align} \label{L_block}
L = -E_8\oplus -E_8\oplus \si\oplus\si\oplus\si,
\end{align}
where $E_8$ is the Cartan matrix of the Lie group $E_8$, and
$\si=\big(\begin{smallmatrix} 0&1 \\ 1&0 \end{smallmatrix} \big)$.

In the case of 10-d heterotic string, the 16 modes of the
left-movers construct an even self-dual lattice of signature
$(16,0)$. There are two such lattices, one of which is
$\Ga_8\times\Ga_8$ with $\Ga_8$ being the root lattice of $E_8$.
Then, as we construct the 10-d heterotic action from the 7-d
action, it is natural to eliminate the block
$\Si=\si\oplus\si\oplus\si$ of signature $(3,3)$ in
eq.~\eqref{L_block} along with the elements in the action that
couple to $\Si$.

In fact, $L^+$, the intersection matrix of the three self-dual
two-forms, $J$, $\frac{1}{2}(\Om+\bOm)$ and  $\frac{1}{2i}(\Om-\bOm)$,
has the signature of $(0,3)$. 
Therefore, $L^+$ forms the sub-block of $\Si$ whose signature is $(3,3)$,
so that we can eliminate $L^+$ along with the $Z_\al^I$ and $\tZ^I$ which couple to it.
Since $Z_\al^I=\p_\al Y^I - c_\al^I$ is the supersymmetric extension of
right-movers $\p_\al Y_+^I$, the elimination of these terms naturally
give the 10-d theory and we are left with scalar degrees of freedom
representing $E_8 \times E_8$ current algebra.

Consequently, keeping only the terms including the sub-block
$L=-E_8\oplus -E_8$ of $L^-$ (and similarly for $M^-$), 
its signature being $(16,0)$, the proposed 10-d action is written as
\begin{align} \label{10d_action}
\cL_1 &= -\sqrt{-G}\sqrt{1 + \frac{\tY M\tY}{Gu^2}
    +\biggl(\frac{\tY L\tY}{2Gu^2}\biggr)^2}, \nn \\
\cL_2 &= -\frac{\tY L\p_\al Y u^\al}{2u^2}, \nn \\
\cL_3 &= -\ep^{\al\be} \bth\Ga_m\p_\al\th\p_\be X^m \quad (m=1,\cdots,10), \nn \\
\cL_4 &= 0.
\end{align}
Here, $\cL_4=0$ because it has contained only $L^+$ terms in 7-d.
Also note that we have added three noncompact coordinates $X^m$ which
enter the induced metric $G_{\al\be}$ in $\cL_1$ as well as in
$\cL_3$. The above 10-d action has correct field contents of the
heterotic string. $Y^I$ are the 16 scalars and $X^m$ are coordinates
of 10 dimensions while $\th$ is a Majorana-Weyl spinor of 10-d with
16 real degrees of freedom.

Note that in the above 10-d action the condition $ML^{-1}M=L$ in 
$(10-n)$ dimension~\cite{cherkis} reduces to $M=-L$. This is
realized from the M5-brane point of view as
$L_{IJ}=\int_{K3}b_I\wedge b_J$ and $M_{IJ}=\int_{K3}b_I\wedge *b_J$
where $b_I$ are taken to be all anti-self-dual when they construct
$(16,0)$ block, as argued above.

The condition $M=-L$ is also well understood in terms of the 10-d
heterotic string. This can be seen first from the moduli space of the theory. 
In $(10-n)$ dimension matrix $M$ characterizes the Narain
moduli space, up to the T-duality group $O(16+n,n;\mathbb Z)$~\cite{cherkis},
\begin{align}
O(16+n,n)/(O(16+n)\times O(n)).
\end{align}
Then, in 10-d ($n=0$) the moduli space M characterizes becomes a 0-dimensional space
so that $M$ is a matrix of fixed components for corresponding $L$.
The concrete form $M=-L$ can be justified
by the projections $\cP_\pm = \frac{1}{2}(1\pm L^{-1}M)$,
for $(L^{-1}M)^2=1$, which projects out right- and left-moving $Y^I$.
Then, in 10-d $\cP_+ = 0$ is needed, since there are no right movers
in compact dimension. As a result, we have $M=-L$.

The equation of motion of $Y^I$ in terms of $P_-$ 
along with the identification of $Y^I$ as chiral bosons (left-movers)
in 10-d action can be given by the PST gauge as follows. 
First we can obtain the compact part of the 10-d action in Polyakov type
in a similar way to the 7-d bosonic string described in ref.~\cite{cherkis}:
\begin{align} \label{pst}
\cL_C = - \frac{1}{2u^2} \tY L(\tY + \p Y\cdot u),
\end{align}
where $M=-L$ has been used and $u_\al=\p_\al a$. This compact part
(as well as the full action) is invariant under the two kinds of PST
gauge transformations
\begin{align} \label{pst_gauge}
\de Y &= \phi\frac{\cP_-\p_+ Y}{\p_+ a},
\quad \de a=\phi, \nn \\
\de Y &= f(a), \quad \de a=0,
\end{align}
where the lattice indices $I$ of $Y^I$ are suppressed. Then fixing
the second gauge in eq.~\eqref{pst_gauge}, similarly to the cases
introduced in ref.~\cite{pasti2}, we can identify the $Y^I$ in the
action as left-movers. As we write down the action in
eq.~\eqref{pst} as
\begin{align}
\cL_C = -\frac{1}{2u^2} u^\al F^*_\al L (F^*_\be + F_\be) u^\be,
\end{align}
where $F_\al=\p_\al Y$, $F^*_\al=\ep_\al\!^\be \p_\be Y$, the e.o.m.
for $Y^I$ becomes
\begin{align} \label{eom}
\ep^{\al\be}\p_\al \Big(\frac{u_\be \cF_\ga u^\ga}{u^2}\Big) =
\ep^{\al\be}\p_\al \Big(\frac{\cF}{u_+}\Big)u_\be =0,
\end{align}
where $\cF=\cF_\al=F^*_\al+F_\al=F_0+F_1=\p_+ Y$. The general
solution of the e.o.m. \eqref{eom} is
\begin{align}
\cF =g(a)u_+,
\end{align}
where $g(a)$ is an arbitrary scalar function of $a$. However, we
find that this solution is equal to the pure gauge $\de \cF$ for the
2nd gauge symmetry of \eqref{pst_gauge}:
\begin{align}
\de \cF =\p_+\de Y = \p_+ f(a) = g(a)u_+,
\end{align}
where $g(a)=f'(a)$. Therefore, we can pick up the solution
\begin{align}
\cF = F^*_\al+F_\al = \cP_-\p_+Y = 0,
\end{align}
as a gauge fixed e.o.m., $F_\al$ being anti-self-dual and $Y^I$
left-movers. (For $L=M$, we get the self-dual $F_\al$ and right
moving $Y^I$.)

Another clue for $M=-L$ is the kappa invariance of the 10d action,
which is not achieved unless $M=-L$ as will be shown below and in
the appendix. Thus the condition $M=-L$ originated from the M5-brane
side is consistent with 10-d heterotic string. We emphasize that at
the classical level we don't have to impose $L$ to be intersection
matrix of the even self-dual lattice with signature (16,0). Such
condition would arise from the absence of the world-sheet anomaly,
which would arise at one-loop. It would be interesting to work out
this explicitly.

Now, let us check the kappa symmetry of the above 10-d action to see
it is the correct one. We can infer, from the kappa variation of the
fields of M5-brane action, the kappa variations of the 10-d fields
\begin{align}
\de\bth = \bar{\ka}(1-\Ga), \quad \de X^m = -\de\bth\Ga^m\th, \quad
\de Y^I = 0.
\end{align}
Note $\de Y^I=0$ which is satisfied in the absence of the Yang-Mills
background $A_m^I$. That is, the kappa variations of 16 left-movers
in compact dimensions vanish. Therefore, we can consider the
variations of $X^m$ and $\th$ of noncompact 10-d only.

From above variations of $X^m$ and $\th$ we can deduce the variation
of the pullback vielbein $\Pi_\al^\hm$. Since here we take the 10-d
supergravity background to be flat,
\begin{align}
\Pi_\al^\hm =\Pi_\al^m =\p_\al X^m-\bth\Ga^m\p_\al\th,
\end{align}
whose kappa variation becomes
\begin{align}
\de\Pi^m_\al=-2\de\bth\Ga^m\p_\al\th.
\end{align}
Then, we get the variations of induced metric $G_{\al\be}$ and the
related quantities as
\begin{align}
\de G_{\al\be} &= -2\de\bth\ga_{\{\al}\p_{\be\}}\th, \nn \\
\de G^{\al\be} &= -G^{\al\ga}\de G_{\ga\de}G^{\de\be}
    = 2\de\bth\ga^{\{\al}\p^{\be\}}\th, \nn \\
\de G &= \frac{1}{2}\ep^{\al\be}\ep^{\ga\de}
    \de(G_{\al\ga}G_{\be\de})
    = -4G\de\bth\ga^\al\p_\al\th, \nn \\
\de u^2 &= \de G^{\al\be}u_\al u_\be
    = 4\de\bth\ga^\al\p^\be\th u_\al u_\be,
\end{align}
where $\ga_\al=\Ga_m\Pi^m_\al$.

We can prove the kappa symmetry of the action by using the variations above.
The main scheme is similar to the one introduced in ref.~\cite{park} 
for the case of M5-brane action. First, define $U^\al$ and $T^\al$ as
\begin{align}
\de \cL_1 &= \frac{2}{\cL_1}\de\bth U^\al\p_\al\th, \nn \\
\de \cL_2 + \de \cL_3 &= 2\de\bth T^\al\p_\al\th.
\end{align}
Evaluation of $\de\cL_1$, $\de\cL_2$, and $\de\cL_3$ gives
\begin{align} \label{UT}
U^\al &= G\ga^\al + \frac{\tY M\tY}{(u^2)^2}\ga^\be u_\be u^\al
    + \frac{(\tY L\tY)^2}{4G(u^2)^3}
    (2\ga^\be u^\al - \ga^\al u^\be)u_\be, \nn \\
T^\al &= -\bga\ga^\al
    + \frac{\tY L\tY}{2(u^2)^2} \ep_{\be\ga}
    \ga^{\{\be}G^{\de\}\al} u^\ga u_\de,
\end{align}
where $\bga=\ga_{01}$. Then, we can take a quantity
\begin{align}
\rho = \bga - \frac{\tY L\tY}{2Gu^2} \bga,
\end{align}
which satisfies, for $M=-L$,
\begin{align} \label{pf1}
U^\al = \rho T^\al, \quad \rho^2 = (\cL_1)^2.
\end{align}
Eq.~\eqref{pf1} above implies that the action is invariant under the
kappa symmetry. To see this, first note that we can take $\de \th =
\bka(1-\Ga)$ with $\Ga=\frac{\rho}{\cL_1}$ because $\Ga^2=1$ from
$\rho^2 = (\cL_1)^2$. Then, by using $U^\al = \rho T^\al$, we find
that
\begin{align} \label{pf2}
\de \cL &= 2\de\bth \biggl(\frac{U^\al}{\cL_1}
    + T^\al\biggr) \p_\al\th \nn \\
&= 2\bka(1 - \Ga)(1 + \Ga) T^\al\p_\al\th \nn \\
&=0.
\end{align}
The details of the proof, including the evaluation of $U^\al$,
$T^\al$, and $\rho$ along with the check of eq.~\eqref{pf1}, are
given in the appendix.

The requirement of the kappa invariance restricts the supersymmetry
of the heterotic action to be $\mathcal{N} =1$. For example, let us
check the possibility of kappa invariance for the case of
$\mathcal{N} =2$ supersymmetry such that $\th=\th_1+\th_2$, $\th_2$
and $\th_2$ being 16-component Majorana-Weyl spinors. Then, $\cL_1$
and $\cL_2$ have the same form as in the eq.~\eqref{10d_action} with
$\bth\Ga^m\p_\al\th$ replaced by $\bth_A\Ga^m\p_\al\th_A$ for
$A=1,2$. But $\cL_3(=\cL_{WZ})$ is written as
\begin{align}
\cL_3= -\ep^{\al\be}(\bth_1\Ga_m\p_\al\th_1 -
\bth_2\Ga_m\p_\al\th_2)\p_\be X^m
    + \ep^{\al\be}\bth_1\Ga_m\p_\al\th_1 \bth_2\Ga^m\p_\be\th_2,
\end{align}
which is the same as $\cL_{WZ}$ in Type II theories. Then, under
kappa transformations, all terms in $\de \cL_1$, $\de \cL_2$ and $\de
\cL_3$ are written in terms of $\de\bth_1$ and $\de\bth_2$
separately:
\begin{align} \label{het}
\de \cL_1 &= \frac{2}{\cL_1}(\de\bth_1 U^\al\p_\al\th_1 
	+ \de\bth_2 U^\al\p_\al\th_2), \nn \\
\de \cL_2 &= 2(\de\bth_1 R^\al \p_\al\th_1 + \de\bth_2 R^\al \p_\al\th_2), \nn \\
\de \cL_3 &= 2(\de\bth_1 S^\al \p_\al\th_1 - \de\bth_2
S^\al\p_\al\th_2),
\end{align}
where $U^\al$ and $T^\al=R^\al +S^\al$ ($S^\al=-\bga\ga^\al$) are
those of eq.~\eqref{UT} with $\bth\Ga^m\p_\al\th$ replaced by
$\bth_A\Ga^m\p_\al\th_A$. Here, for $\de L$ to vanish, $\de\bth_1$
and $\de\bth_2$ terms should vanish separately. However, this is not
possible due to the sign difference between the two terms in $\de
\cL_3$. That is, referring to eq.~\eqref{pf1} and \eqref{pf2}, we
need to have for kappa invariance
\begin{align}
U^\al = \rho_1 (R^\al + S^\al), \quad U^\al = \rho_2 (R^\al - S^\al),
 \quad \rho_1^2=\rho_2^2=(\cL_1)^2,
\end{align}
which cannot be satisfied simultaneously. Thus the action is not
kappa invariant.

In contrast to the heterotic case, Type II theories have kappa
symmetry with $\th=\th_1+\th_2$. A Type II action in flat background
can be given from the above heterotic action by taking $Y^I=0$ and
$\th=\th_1+\th_2$:
\begin{align}
\cL_1 &= -\sqrt{-G}, \nn \\
\cL_2 &= 0, \nn \\
\cL_3 &= -\ep^{\al\be}(\bth_1\Ga_m\p_\al\th_1 -
\bth_1\Ga_m\p_\al\th_1)\p_\be X^m
    + \ep^{\al\be}\bth_1\Ga_m\p_\al\th_1 \bth_2\Ga^m\p_\be\th_2.
\end{align}
Here $\cL_2=0$ is critical for the kappa invariance of the action.
The action transforms as
\begin{align}
\de \cL_1 &= \frac{2}{\cL_1}(\de\bth_1 U^\al\p_\al\th_1 + 
	\de\bth_2 U^\al\p_\al\th_2), \nn \\
\de \cL_3 &= 2(\de\bth_1 S^\al\p_\al\th_1 - \de\bth_2
S^\al\p_\al\th_2),
\end{align}
where $U^\al$ is different from those of eq. \eqref{het}. Then it is
possible to take
\begin{align}
U^\al=\rho S^\al, \quad \rho^2=(\cL_1)^2,
\end{align}
which results in, for $\Ga=\frac{\rho}{L_1}$,
\begin{align}
\de \cL = 2\de\bth_1(1+\Ga)S^\al\p_\al\th_1 +
2\de\bth_2(1-\Ga)S^\al\p_\al\th_2 = 0,
\end{align}
for the $\th_A$ variations
\begin{align}
\de\bth_1 = \bka_1(1-\Ga), \quad \de\bth_2 = \bka_2(1+\Ga).
\end{align}
Here the $\de\th_1$ and $\de\th_2$ terms vanish separately.

\section{Discussions}
In this paper we propose the supersymmetric heterotic string action
motivated by the dimensional reduction of M5-brane wrapping on K3.
And we explicitly prove the kappa symmetry of the resulting
heterotic action. The novelty is the additional scalars realizing
the chiral current algebra. It would be interesting to explicitly
verify underlying Kac-Moody algebras. Also one should consider the
generalization of the heterotic action in an arbitrary supergravity
background. This construction can also be viewed as D1-D9 system in
the Type IIB theory with treating D9s as backgrounds. In Type IIB,
there is also D9-D5 system, which is supersymmetric. Its DBI type
action is unknown. It would be interesting to work out D9-D5 action
and we hope that the current construction can shed some light on
that problem.

\subsection*{Acknowledgments}

J.P. is  supported in part by the KOSEF SRC Program
through CQUeST at Sogang University, by KOSEF Grant R01-2008-000-20370-0
and by  the Stanford Institute for Theoretical Physics.

\medskip
\section*{Appendix}
\appendix

\section{Proof of the Kappa Symmetry in 10-d Action} \label{sec:appendix}

\subsection{Useful relations}

For $\ga_\al =\Ga_m \Pi^m_\al$ and $\bga =\ga_{01}$,
\begin{alignat}{2} \label{usf}
\bga\ga_{\al} &= -G\ep_{\al\be}\ga^\be,
    &\quad \bga\ga_{\al\be} &= -G\ep_{\al\be}, \nn \\
\ga_\al\ga_\be &= \ga_{\al\be} +G_{\al\be},
    & \ga_{\al\be}\ga_\ga &= \ga_{[\al}G_{\be]\ga}.
\end{alignat}
The generalization to a worldvolume of arbitrary dimension is, for
$\bga =\ga_{1\cdots D}$,
\begin{align}
\bga\ga_{\mu_{k+1}\cdots\mu_D} &=
(-)^{\frac{(k+1)(k+2)}{2}}\frac{G}{k!}\ep_{\mu_1\cdots\mu_D}\ga^{\mu_1\cdots\mu_k}, \nn \\
\ga_{\mu_1\cdots\mu_m}\ga_{\nu_1\cdots\nu_n} &=
\sum_{k=0}^{\min(m,n)}C_k^{mn}\ga_{\mu_1\cdots\mu_{m-k}\nu_1\cdots\nu_{n-k}}
    G_{\mu_{m-k+1}\nu_{n-k+1}}\cdots G_{\mu_m\nu_n},
\end{align}
where $C_k^{mn}\equiv (-)^{kn+k(k+1)/2}k!
\bigl(\begin{smallmatrix}m\\k\end{smallmatrix}\bigr)
\bigl(\begin{smallmatrix}n\\k\end{smallmatrix}\bigr)$ and $\mu$'s and
$\nu$'s in the RHS are antisymmetrized separately.

\subsection{Evaluation of $U^\al$}
First, let us evaluate $\de (\cL_1)^2$ to get $U^\al$ from $\de
(\cL_1)^2 =2\cL_1\de \cL_1 =4\de\bth U^\al\p_\al\th$. Since $\de
\tY^I=0$ from $\de Y^I=0$, we can consider $\de G$ and $\de u^2$
above only:
\begin{align}
\de (\cL_1)^2 &= -\de G - (\tY M\tY)\de\biggl(\frac{1}{u^2}\biggr)
    -\frac{(\tY L\tY)^2}{4}\de\biggl(\frac{1}{G(u^2)^2}\biggr) \nn \\
    &= 4\de\bth \biggl(
    G\ga^\al + \frac{\tY M\tY}{(u^2)^2}\ga^\be u_\be u^\al
    + \frac{(\tY L\tY)^2}{4G(u^2)^3}
    (2\ga^\be u^\al - \ga^\al u^\be)u_\be
    \biggr) \p_\al\th,
\end{align}
which gives
\begin{align}
U^\al = G\ga^\al + \frac{\tY M\tY}{(u^2)^2}\ga^\be u_\be u^\al
    + \frac{(\tY L\tY)^2}{4G(u^2)^3}
    (2\ga^\be u^\al - \ga^\al u^\be)u_\be.
\end{align}

\subsection{Evaluation of $T^\al$}
First let us evaluate $\de \cL_2$:
\begin{align}
\de \cL_2 &= -\frac{\tY L\p_\be Y u_\al}{2u^2}(\de G^{\be\al})
    -\frac{\tY L\p_\be Y u^\be}{2}
    \de\biggl(\frac{1}{u^2}\biggr) \nn \\
&= \de\bth\biggl(
    -\frac{ \tY L\p_\be Y u_\al \ga^{\{\be}\p^{\al\}}\th }{u^2}
    + \frac{ 2\tY L\p_\be Y u^\be \ga^\ga \p^\al\th u_\ga u_\al }{(u^2)^2}
    \biggr) \nn \\
&= \de\bth\frac{\tY L}{(u^2)^2}\biggl(
    - \p_\be Y \ga^{\{\be} G^{\ga\}\al} u_\ga u^2
    + 2\p_\be Y u^\be \ga^\ga u_\ga u^\al
    \biggr)\p_\al\th.
\end{align}
This can be simplified more by rewriting the second term in the
parenthesis as
\begin{align}
2\p_\be Y u^\be u_\ga \ga^\ga u^\al
&= \p_\ga Y u^\ga u_\be(\ga^\be u^\al + \ga^\de u_\de G^{\be\al}) \nn \\
&= (\p_\be Y u^2 + \tY\ep_{\be\ga}u^\ga)
    (\ga^\be u^\al + \ga^\de u_\de G^{\be\al}) \nn \\
&= \p_\be Y \ga^{\{\be} G^{\ga\}\al} u_\ga u^2 +
    \tY\ep_{\be\ga}u^\ga (\ga^\be u^\al + \ga^\de u_\de G^{\be\al}),
\end{align}
where we have used
\begin{align}
\p_\ga Y u^\ga u_\be = \p_\be Y u^2 + \tY\ep_{\be\ga}u^\ga.
\end{align}
Then this cancels out the first term of $\de \cL_2$ giving
\begin{align}
\de \cL_2 &= \de\bth\frac{\tY L\tY}{(u^2)^2} \ep_{\be\ga}u^\ga
    (\ga^\be u^\al + \ga^\de u_\de G^{\be\al})\p_\al\th \nn \\
&= 2\de\bth \biggl( \frac{\tY L\tY}{2(u^2)^2} \ep_{\be\ga}
    \ga^{\{\be}G^{\de\}\al} u^\ga u_\de \biggr) \p_\al\th.
\end{align}

Now let us evaluate $\de \cL_3$,
\begin{align}
\de \cL_3 &= -\ep^{\al\be} \de(\bth\Ga_m\p_\al\th) \p_\be X^m
    -\ep^{\al\be} (\bth\Ga_m\p_\al\th) \p_\be \de X^m.
\end{align}
Here the first term becomes
\begin{align}
-\ep^{\al\be} \de(\bth\Ga_m\p_\al\th) \p_\be X^m &= -\ep^{\al\be}
    (\de\bth\Ga_m\p_\al\th + \bth\Ga_m\p_\al\de\th) \p_\be X^m \nn \\
&= -\ep^{\al\be}
    (\de\bth\Ga_m\p_\al\th - \p_\al\de\bth\Ga_m\th) \p_\be X^m \nn \\
&= -\ep^{\al\be} (2\de\bth\Ga_m\p_\al\th) \p_\be X^m
    - \underbrace{ \ep^{\al\be} \p_\al(\de\bth\Ga_m\th) \p_\be X^m }_
    { \text{total derivative} },
\end{align}
where the second equality is realized since $\th$ is Majorana. Next,
the second term becomes
\begin{align}
-\ep^{\al\be} (\bth\Ga_m\p_\al\th) \p_\be \de X^m
&= \ep^{\al\be} \p_\al \de X_m (\bth\Ga^m\p_\be\th) \nn \\
&= \ep^{\al\be} \p_\al (-\de\bth\Ga_m\th) (\bth\Ga^m\p_\be\th) \nn \\
&= \ep^{\al\be} (\de\bth\Ga_m\th)(\p_\al\bth\Ga^m\p_\be\th)
    + \underbrace{ \ep^{\al\be} \p_\al (-\de\bth\Ga_m\th\bth)\Ga^m\p_\be\th }_
    { \text{total derivative} }.
\end{align}
Moreover, we can use a spinor identity~\cite{green}, which is realized for
Majorana-Weyl spinors in 10-d, to change the last line as
\begin{align}
\ep^{\al\be} (\de\bth\Ga_m\th)(\p_\al\bth\Ga^m\p_\be\th) =
2\ep^{\al\be} (\de\bth\Ga_m\p_\al\th)(\bth\Ga^m\p_\be\th).
\end{align}
Finally, collecting the two terms of $\de \cL_3$ again,
\begin{align}
\de \cL_3 &= -2\ep^{\al\be} (\de\bth\Ga_m\p_\al\th)
    (\p_\be X^m - \bth\Ga^m\p_\be\th) \nn \\
&= -2\ep^{\al\be} (\de\bth\Ga_m\p_\al\th) \Pi_\be^m \nn \\
&= 2\de\bth (-\ep^{\al\be}\Ga_m\Pi_\be^m) \p_\al\th \nn \\
&= 2\de\bth (-\ep^{\al\be}\ga_\be) \p_\al\th \nn \\
&= 2\de\bth (-\bga\ga^\al) \p_\al\th.
\end{align}

In result, we get $T^\al$ from $\de \cL_2 +\de \cL_3$ as
\begin{align}
T^\al = -\bga\ga^\al
    + \frac{\tY L\tY}{2(u^2)^2} \ep_{\be\ga}
    \ga^{\{\be}G^{\de\}\al} u^\ga u_\de.
\end{align}

\subsection{Determination of $\rho$}
Equipped with $U^\al$ and $T^\al$, we now find an appropriate $\rho$
with $\rho^2=(\cL_1)^2$ and show $U^\al =\rho T^\al$ to complete the
proof. First, we take $\rho$ as
\begin{align}
\rho = \bga - \frac{\tY L\tY}{2Gu^2} \bga.
\end{align}
Then $\rho^2$ becomes
\begin{align}
\rho^2 &= \bga^2 - \frac{\tY L\tY}{Gu^2}\bga^2
    + \frac{(\tY L\tY)^2}{4G^2(u^2)^2}\bga^2 \nn \\
&= -G - \frac{\tY M\tY}{u^2}
    - \frac{(\tY L\tY)^2}{4G(u^2)^2} \nn \\
&= (\cL_1)^2,
\end{align}
where $\bga^2=-G$ and $M=-L$ have been used.

\subsection{Proof of $U^\al =\rho T^\al$}
To show $U^\al =\rho T^\al$, we decompose $U^\al$, $T^\al$ and $\rho$
according to the power of $\tY^I$ as
\begin{align}
U_0^\al &= G\ga^\al, \nn \\
U_1^\al &= 0, \nn \\
U_2^\al &= \frac{\tY M\tY}{(u^2)^2}\ga^\be u_\be u^\al, \nn \\
U_3^\al &= 0, \nn \\
U_4^\al &= \frac{(\tY L\tY)^2}{4G(u^2)^3}
    (2\ga^\be u^\al - \ga^\al u^\be)u_\be,
\end{align}
\begin{align}
T_0^\al &= -\bga\ga^\al, \nn \\
T_1^\al &= 0, \nn \\
T_2^\al &= \frac{\tY L\tY}{2(u^2)^2}
    \ep_{\be\ga}\ga^{\{\be}G^{\de\}\al} u^\ga u_\de,
\end{align}
\begin{align}
\rho_0 &= \bga, \nn \\
\rho_1 &= 0, \nn \\
\rho_2 &= -\frac{\tY L\tY}{2Gu^2}\bga,
\end{align}
where the subscripts of $U$, $T$ and $\rho$ denote the powers of
$\tY$. Then $U^\al = \rho T^\al$ becomes
\begin{align}
U_0^\al &= \rho_0 T_0^\al, \nn \\
U_1^\al &= \rho_0 T_1^\al + \rho_1 T_0^\al, \nn \\
U_2^\al &= \rho_0 T_2^\al + \rho_1 T_1^\al + \rho_2 T_0^\al, \nn \\
U_3^\al &= \rho_1 T_2^\al + \rho_2 T_1^\al,\nn  \\
U_4^\al &= \rho_2 T_2^\al.
\end{align}

Then we can complete the proof of kappa symmetry by checking above
relations. First $U_1^\al = \rho_0 T_1^\al + \rho_1 T_0^\al$ and
$U_3^\al = \rho_1 T_2^\al + \rho_2 T_1^\al$ are checked easily, since
$T_1^\al=\rho_1=0$ and $U_1^\al=U_3^\al=0$.

$U_0^\al = \rho_0 T_1^\al + \rho_1 T_0^\al$ is checked as
\begin{align}
\rho_0 T_0^\al = -\bga^2\ga^\al = G\ga^\al = U_0^\al,
\end{align}
where we have used $\bga^2=-G$.

Next see how $U_2^\al = \rho_0 T_2^\al + \rho_1 T_1^\al + \rho_2
T_0^\al$ is realized. First $\rho_0 T_2^\al$ is evaluated as,
omitting the factor $\frac{\tY L\tY}{2(u^2)^2} u^\ga u_\de$,
\begin{align}
\rho_0 T_2^\al
&\sim \bga \ep_{\be\ga} \ga^{\{\be}G^{\de\}\al} \nn \\
&= \ga_{\be\ga} \ga^{\{\be}G^{\de\}\al} \nn \\
&= \ga_{\be\ga} \ga_\ep G^{\ep\{\be}G^{\de\}\al} \nn \\
&= \ga_{[\be} G_{\ga]\ep} G^{\ep\{\be}G^{\de\}\al} \nn \\
&= \ga_{[\be} \de_{\ga]}^{\{\be}G^{\de\}\al} \nn \\
&= (\ga_\ga-2\ga_\ga)G^{\de\al}
    + \ga^\al \de_\ga^\de - \ga_\ga G^{\de\al} \nn \\
&= -2\ga_\ga G^{\de\al} + \de_\ga^\de \ga^\al,
\end{align}
where we used $\bga\ep_{\be\ga}=\ga_{\be\ga}$ and
$\ga_{\be\ga}\ga_\ep=\ga_{[\be} G_{\ga]\ep}$. Then,
\begin{align}
\rho_0 T_2^\al &= \frac{\tY L\tY}{2(u^2)^2} u^\ga u_\de
    (-2\ga_\ga G^{\de\al} + \de_\ga^\de \ga^\al) \nn \\
&= \frac{\tY L\tY}{2(u^2)^2}
    (-2\ga^\be u^\al + \ga^\al u^\be)u_\be.
\end{align}
Secondly, $\rho_1 T_1^\al=0$ from $\rho_1=0$. Lastly,
\begin{align}
\rho_2T_0^\al
&= \frac{\tY L\tY}{2Gu^2}\bga\bga\ga^\al \nn \\
&= -\frac{\tY L\tY}{2(u^2)^2}\ga^\al u^\be u_\be,
\end{align}
where $\bga^2=-G$ has been used. Then using $M=-L$,
\begin{align}
\rho_0 T_2^\al + \rho_1 T_1^\al + \rho_2 T_0^\al
&= \frac{\tY M\tY}{(u^2)^2}\ga^\be u_\be u^\al \nn \\
&= U_2^\al.
\end{align}

Finally, $U_4^\al = \rho_2 T_2^\al$ is proved as
\begin{align}
\rho_2 T_2^\al
&= -\frac{\tY L\tY}{2Gu^2} \rho_0 T_2^\al \nn \\
&= \frac{(\tY L\tY)^2}{4G(u^2)^3}
    (2\ga^\be u^\al - \ga^\al u^\be)u_\be \nn \\
&= U_4^\al.
\end{align}
Thus we have proved the kappa symmetry of 10-d heterotic action.

\end{document}